\documentclass[superscriptaddress,twocolumn]{revtex4}
\usepackage{float} 
\usepackage{booktabs}
\usepackage{mathrsfs}
\usepackage{amssymb}
\usepackage{amsmath}
\usepackage{graphicx}
\usepackage{epsfig,graphicx,times}
\usepackage{color}
\usepackage{subfigure}
\usepackage{setspace}
\usepackage{epstopdf}
\usepackage[T1]{fontenc}
\usepackage{bm}
\usepackage{upgreek}
\usepackage{hyperref}
\hypersetup
{
hypertex=true,
colorlinks=true,
linkcolor=blue,
filecolor=blue,
urlcolor=blue,
citecolor=blue,
breaklinks=true
}
\begin{document}
\title{Approaching energy quantum limit detection of microwave photons with Josephson Junctions}
\author{Y. Q. Chai}
\affiliation{Information Quantum Technology Laboratory, International Cooperation Research Center of China Communication and Sensor Networks for Modern Transportation, School of Information Science and Technology, Southwest Jiaotong University, Chengdu 610031, China}

\author{S. N. Wang}
\affiliation{Information Quantum Technology Laboratory, International Cooperation Research Center of China Communication and Sensor Networks for Modern Transportation, School of Information Science and Technology, Southwest Jiaotong University, Chengdu 610031, China}

\author{P. H. OuYang}
\affiliation{Information Quantum Technology Laboratory, International Cooperation Research Center of China Communication and Sensor Networks for Modern Transportation, School of Information Science and Technology, Southwest Jiaotong University, Chengdu 610031, China}

\author{L. F. Wei \footnote{Corresponding author.  lfwei@swjtu.edu.cn }}
\affiliation{Information Quantum Technology Laboratory, International Cooperation Research Center of China Communication and Sensor Networks for Modern Transportation, School of Information Science and Technology, Southwest Jiaotong University, Chengdu 610031, China}

\date{\today}
\begin{abstract}
Single-photon detection is an energy quantum limit detection (EQLD) of a significantly weak electromagnetic wave. Given the sensitivity of the conventional electromagnetic induction microwave receiver working at room-temperature is very limited, due to the significantly strong thermal noise, here we analyze the possibility of approaching the EQLD of the weak microwave signal by using a current-biased Josephson Junction (CBJJ) detector. By numerically simulating the dynamics for the phase particle of the CBJJ, we propose an approach to describe the discriminability between the probabilistically escaped events of the phase particle with and without the microwave current driving, by measuring the minimum $d_{\rm KC}$-index. We predicate that, the experimentally demonstrated CBJJ detectors possess the ability to resolve about a dozen photons. The feasibility of the desired EQLD of microwave signal by using the CBJJ detector is also discussed. 
\end{abstract}
\maketitle
\section{Introduction}
Microwave receiver is a system to sensitively detect the microwave signal in noisy environments. It is well-known that one of the biggest obstacles to improve the detection sensitivity is the thermal noise; e.g., at 290 K room temperature,  the sensitivity is limited as ${\rm 4\times10^{-18}~W/\sqrt{Hz}}$~\cite{Thermal}, which is far from the energy quantum limit detection (EQLD) at the single-photon level~\cite{photon1,photon2,photon3,photon4,photon5,photon6,photon7}. Although the sensitivity can be improved by developing various techniques, such as phase lock amplifications~\cite{LNA}, multi-branch receptions~\cite{RE1}, outlier detection~\cite{RE2}, and digital-to-analog conversions~\cite{RE6}, etc., the reachable sensitivity is still very limited~\cite{RE3,RE4,RE5}. Therefore, the development of the higher sensitivity microwave receiver is still a great challenge, specifically for solid-state quantum computing~\cite{quantum1,quantum2,quantum3,quantum4}, microwave quantum sensing~\cite{Illumination}, and also for the sensitive detection of either gravitational waves in GHz band~\cite{GW1,GW2} or the axion dark matters with $\mu$eV-order masses~\cite{Axion1,Axion2}, etc. 

Physically, the energy of a microwave photon is at $\sim 10^{-24}$~J level, which is 4–5 orders of magnitude lower than the energy of a single photon at the optical band. This implies that the microwave single-photon detector should work in the mK-order temperature regime for the effectively suppressing the relevant thermal noise. For example, if a superconducting microwave receiver is worked at 10 mK, its realizable detection sensitivity could be estimated as $\sim 10^{-23}~{\rm W/\sqrt{Hz}}$~\cite{Thermal}, which is theoretically improved the sensitivity by 5–6 orders of magnitude relative to a room-temperature microwave receiver. Therefore, various cryogenic superconducting devices could be served as the natural candidates to implement the sensitive detection of the weak microwave signal for approaching the corresponding EQLD. 

Until now, a series of approaches have been really demonstrated to implement the desired microwave single-photon detection, i.e., at EQLD level. The first one is called resonant detection, which is usually implemented by probing the microwave single-photon induced resonant quantum transition, between the levels of the  artificial atom~\cite{qubit1,quan-dot1,quan-dot2}. The second one is achieved by using the low-noise Josephson parametric amplifier~\cite{MPA1,MPA2} to amplify the single-photon signal for the later detection. However, what the former realization is the narrow band detection, as its detection efficiency is basically determined by the resonantly or near-resonantly excited probabilities of the ground-state artificial atoms, driven by the microwave photon. While, the detection sensitivity demonstrated by the latter one is very limited, due to the influence from the addition noise of the driving source~\cite{Q-control}.
Therefore, fast, highly efficient, and highly sensitive detection of the microwave single photons are still a great challenge for the desirable microwave quantum information processing (such as the superconducting quantum computation~\cite{super1,super2}, microwave quantum radar and the microwave quantum sensing~\cite{quan-measure}), and also the significantly weak microwave response detections of the $\mu$eV-mass axion dark matter~\cite{Axion1,Axion2} and the microwave band gravitational waves~\cite{GW1}, in the stationary strong magnetic fields.

Focusing on such an urgent requirement, in the present work we investigate the possibility of the fast and sensitive detection of microwave photons, specifically with the experimental current-biased Josephson Junction device. Here, the CBJJ is a highly quality Josephson junction biased by a dc-current, whose amplitude is slightly less than its critical current, and the detected weak microwave signal is treated as a modified biased current. Based on the usual RCSJ model~\cite{RCSJ1,RCSJ2} with the noise current driving, we numerically simulate the dynamics of the phase particle driven by the driven microwave current signal, for describing the switching behavior of the junction being transited from the zero-voltage state into the finite voltage one~\cite{JJ1,JJ2,JJ3,Ali}. By comparing the switching behaviors of the CBJJ with and without the driven microwave signal, we quantify its detectability by the measured
Kumar Caroll (KC) index~\cite{d-kc1,d-kc2}, effectively.

The paper is organized as follows. In Sec.~II, based on the usual RCSJ model, we numerically solve the dynamics for a phase particle driven by thermal noise, and then discuss how to implement the weak microwave signal detection by probing the escape effect of the phase particle. A so-called minimum $\min[d_{\rm KC}]$ index is introduced to characterize the detectability of the junction being switched from the zero-voltage state into the finite voltage state. In Sec.~III, we specifically investigate how to detect microwave current pulse signals containing N photons applied to the CBJJ. With the demonstrated CBJJs, the minimum number of the detectable microwave photons are estimated. In Sec.~IV, we summarize our results and discuss the possibility of using the CBJJ device to implement the EQLD of a single microwave photon.
\section{Dynamical simulation for the escapes of a driven phase particle}
\begin{figure}[htbp]
	\centering
	\subfigure[]{	\includegraphics[width=0.45\linewidth]{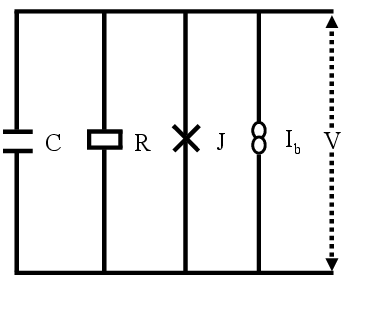}
	}
	\subfigure[]{	\includegraphics[width=0.45\linewidth]{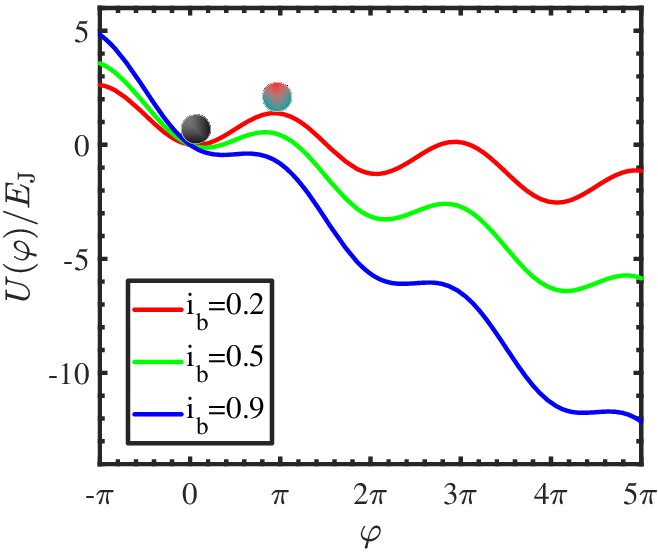}
    }
	\caption{A CBJJ device can be described by: (a) A RCSJ equivalent circuit, where $V$ is the voltage across JJ, $C$, $R$, and $I_{\rm b}$ denote the capacitance, and resistance of the Junction J, respectively. (b) A macroscopic phase particle moving in the washboard potentials, which has been normalized to the Josephson energy $E_{\rm J}=\hbar I_0/2e$, for different biased currents. Here, the red, green, and blue solid lines indicate the washboard potentials generated by different biased currents. The ball represents the states of the CBJJ; the black- and cyan-red balls refer to zero-voltage and the finite voltage states of the CBJJ, respectively.}\label{F1}
\end{figure}

A single Josephson Junction (JJ) without current bias consists simply of two superconducting electrodes sandwiched between a layer of thin insulator, and obeys the following Josephson relation~\cite{Josephson}
\begin{equation}
\begin{aligned}
I_{\rm J}=I_{\rm 0}\sin(\varphi),\,V=\frac{\hbar}{2e}\frac{{\rm d}\varphi}{{\rm d}t}\label{eq:JJ1}
\end{aligned}
\end{equation}
where $I_{\rm J}$, $\varphi$ and $V$ are the Josephson current, macroscopic phase and voltage of the junction, respectively. $I_{\rm 0}$, $\hbar$, and $2e$ are the critical currents (denoting the maximum Cooper-pair tunneling current through the junction), the reduced Planck constant and the Cooper-pair charge, respectively. Under a dc-current $I_{\rm b}$ bias, the device can be described effectively by a capacitive-resistive shunt model (RCSJ model)~\cite{RCSJ1,RCSJ2}, shown in Fig.~\ref{F1}(a),
and thus
\begin{equation}
\left\{
\begin{aligned}
&I_{\rm C}+I_{\rm R}+I_{\rm J}=I_{\rm b}\,,\\
&\frac{\hbar C}{2e}\frac{{\rm d}^2\varphi}{{\rm d}t^2}+\frac{\hbar}{2eR}\frac{{\rm d}\varphi}{{\rm d}t}+I_{\rm 0}\sin(\varphi)=I_{\rm b}\,,\label{eq:RCSJ}
\end{aligned}
\right.
\end{equation}
in the absence of noise. Here, $I_{\rm C}$ and $I_{\rm R}$ represent the currents flowing through the capacitor and resistor, respectively.
Formally, such a current-biased Josephson junction (CBJJ) device can be described as a ``phase particle'' of ``mass'' $m=C(\hbar/2e)^2$ moving in a so-called the washboard potential~\cite{RCSJ3},
\begin{equation}
\begin{aligned}
U(\varphi)=E_{\rm J}\left[1-\cos(\varphi)-i_{\rm b}\varphi\right],\,i_{\rm b}<1\label{eq:potential},
\end{aligned}
\end{equation}
where $i_{\rm b}=I_{\rm b}/I_0$,  and shown schematically in Fig.~\ref{F1}(b). The height of the potential barrier is~\cite{RCSJ4}
\begin{equation}
\begin{aligned}
\Delta U(\varphi)=2E_{\rm J}\left[\sqrt{1-i_{\rm b}^2}-i_{\rm b}\arccos(i_{\rm b})\right]\,.\label{eq:high}
\end{aligned}
\end{equation}
Phonemically, depending on if the phase of the particle is less or larger than the critical phase $\varphi^{*}$, i.e., the trapped particle is at its highest energy value, the CBJJ is in either the superconducting state with $V=0$ (i.e., the color of the ``phase particle" is black) or the finite voltage state with $V\neq 0$ (i.e., the color of the ``phase particle" is cyan-red). Correspondingly, the current flowing through the JJ satisfies the condition; either $i_{\rm b}>1$, or $i_{\rm b}<1$.
Therefore, for a CBJJ with a dc-current $i_{\rm b}$, the slowly added microwave current $\delta i$ increases the energy of the phase particle consequently, until it escapes finally from the potential well. This means that by probing the induced finite voltage state transited from the zero-voltage, i.e., the color of the particle is transited from black to cyan-red, the applied microwave current with the relevant energy can be detected.  
\subsection{Dynamics of the coherently driven phase particle in absence of noise and the voltage-state switch}
Besides the dc-current bias, we first treat simply the applied microwave as an additional weak coherent current,
\begin{equation}
\begin{aligned}
I_{\rm s}=I_{\rm MW}\sin(\omega_{\rm s}t)\,,\label{eq:signal}
\end{aligned}
\end{equation}
with $\omega_{\rm s}$ and $I_{\rm MW}$ are the frequency and amplitude of the applied microwave current, respectively. As a consequence, the RCSJ equation shown in Eq.~\eqref{eq:RCSJ} can be modified as
\begin{equation}
\begin{aligned}
\frac{{\rm d}^2\varphi}{{\rm d}\tau^2}+\beta\frac{{\rm d}\varphi}{{\rm d}\tau}+\sin(\varphi)=i_{\rm b}+i_{\rm s}\,,\label{eq:RCSJ1}
\end{aligned}
\end{equation}
where $\beta=1/RC\omega_{\rm J}$, $i_{\rm s}=I_{\rm s}/I_{\rm 0}$ and $i_{\rm b}=I_{\rm b}/I_{\rm 0}$ are the damping parameter, the normalized bias current, and the weak signal current, respectively. $\omega_{\rm J}=\sqrt{2eI_{\rm 0}/\hbar C}$ is the plasma frequency of the JJ~\cite{RCSJ3}. 
Physically, without such a coherent microwave current, the phase particle damply oscillates in the washboard potential, whose height is defined by the normalized biased dc-current $i_{\rm b}$.
Due to the applied microwave current $i_{\rm s}$, the barrier height of the washboard potential oscillates also with the frequency of $\omega_{\rm s}$. This implies that the phase particle possesses certain probabilities to be escaped out of the well. It is an observed fact that the phase particle is more likely to escape from the potential well if the frequency of the microwave signal is near resonance with JJ's Plasma frequency. However, it is noted that the Plasma frequency of CBJJ is modulated as $\omega_{\rm J}^{*}=(1-i_{\rm b}^2)^{1/4}\omega_{\rm J}$, due to the bias current $i_{\rm b}$. This makes the transition between the CBJJ voltage states extremely sensitive to the microwave signal, if the $\omega_{\rm s}\cong\omega_{\rm J}^{*}$.

To numerically simulate the escape behavior of the driven phase particle, we introduce a critical phase value $\varphi^*$ for the phase variable $\varphi(\tau),\tau=\omega_{\rm J}t$ and define $V=0$ for $\varphi\leq\varphi^*$and $V\neq 0$ for $\varphi>\varphi^*$. By randomly-selected initial phase $\varphi(0)=[-0.1, 0.1]$, the initial voltage $V(0)=d\varphi/d\tau|_{\tau=0}=0$ and also the critical phase $\varphi^*=\pi$, one can numerically solve Eq.~\eqref{eq:RCSJ1} to simulate the possible escape events of the phase particle; once the phase variable
$\varphi(\tau)$ is evolved to the time $\tau^*$ with
\begin{figure}[htbp]
	\centering
	\subfigure[]{\includegraphics[width=0.45\linewidth]{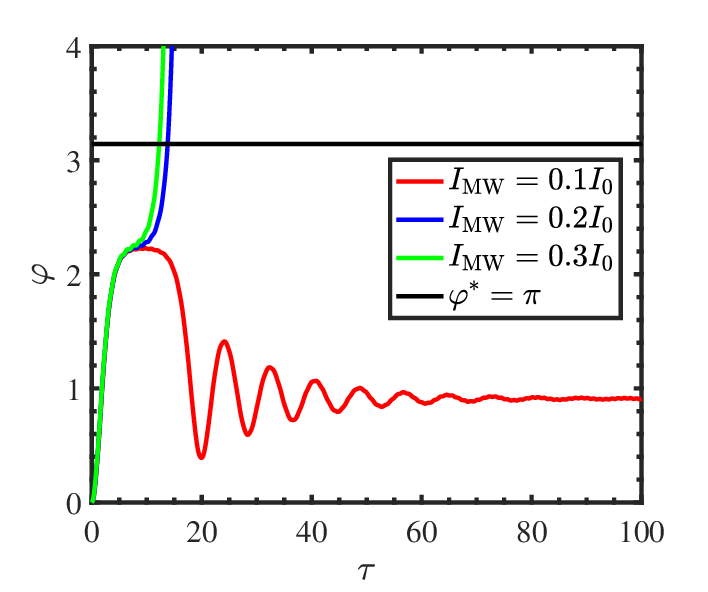}
	}
	\subfigure[]{\includegraphics[width=0.45\linewidth]{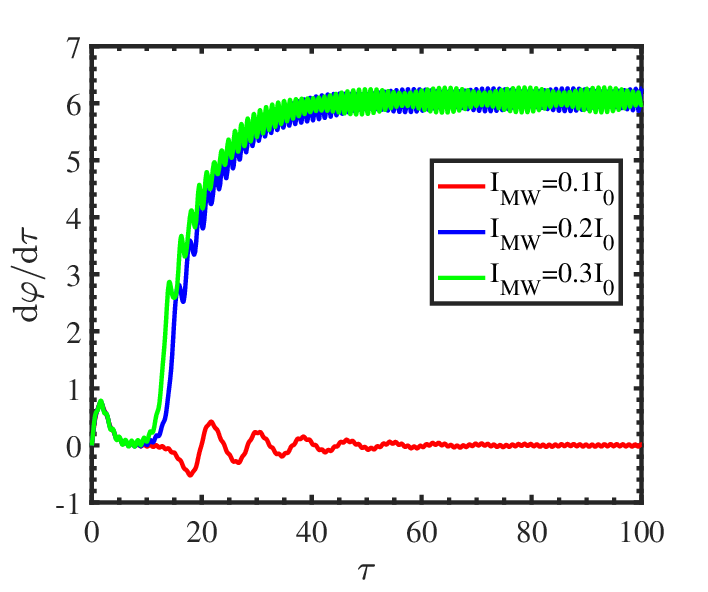}
    }
	\caption{The simulated phase dynamics (a) and voltage changes (b), by numerically solving  Eq.~\eqref{eq:RCSJ1}. Here, the length of the time step is set as $\delta t=0.0001$ and the JJ parameters are set as~\cite{JJ-Al1}: $I_{\rm 0}=8.586~\upmu$A, $R=29~\Omega$, and $C=2700$~fF, respectively. The biased dc-current and the frequency of the microwave signal are set as $i_{\rm b}=0.789$ and $\omega_{\rm s}/2\pi\approx12.25$~GHz, respectively.}\label{F3}
\end{figure}
$\varphi(\tau^*)>\pi$, we get an event of the JJ being switched from the superconducting state into the finite voltage state with $\overline{\dot{\varphi}(\tau)}\neq 0$, and then run another simulation. Repeating such a numerical simulation multiple times. Fig.~\ref{F3} shows these switching behaviors for signals with different amplitudes, wherein the red lines refer to that the JJ still stays at the zero-voltage state and the green and blue lines indicate that the JJ has been switched into the finite voltage states with $\varphi>\varphi^{*}$ and $\overline{\dot{\varphi}(\tau)}\neq 0$.
Therefore, weak microwave signals can be detected by using the CBJJ devices; the larger the bias dc-current $i_{\rm b}$, which yields the value of $i_{\rm b}$ is closer to 1, and thus the smaller amplitude $I_{\rm MW}$ of the signal can be detected, for the given JJ parameters. The stronger the microwave signal corresponds to the larger the measurable voltage.
Ideally, an arbitrarily weak microwave signal can be detected, if the bias current $i_{\rm b}$ flowing through the CBJJ is regulated such that $i_{\rm b}$ is infinitely close to 1.
Obviously, this is impossible, as any actual device is necessarily noisy. 
\subsection{Detectability of the escape events of coherently driven phase particle in the presence of noise: the Kumar-Caroll index}
In practice, the switching of the JJ voltage state is due to not only the microwave weak signal, which is expected to be detected, but also various practically-existing noises, typically such as the thermal-~\cite{thermal} and macroscopic quantum tunneling fluctuations~\cite{MQT,shot,LE}. Below, we investigate specifically how the applied coherent microwave signal yields the escape of the phase particle from the potential well, specifically in the presence of thermal noise. To this end, we solve the following Langevin equation
\begin{equation}
\begin{aligned}
\frac{{\rm d}^2\varphi}{{\rm d}\tau^2}+\beta\frac{{\rm d}\varphi}{{\rm d}\tau}+\sin(\varphi)=i_{\rm b}+i_{\rm n}\,,\label{eq:Langevin}
\end{aligned}
\end{equation}
instead of Eq.~\eqref{eq:RCSJ1}. Above, $i_{\rm n}=I_{\rm n}/I_{\rm 0}$ is the normalized noise current, which satisfies the following statistical distribution feature~\cite{Nyquist}
\begin{equation}
\begin{aligned}
\langle i_{\rm n}\rangle&=0\,,\\
\langle i_{\rm n}(\tau)i_{\rm n}(\tau')\rangle&=4\int_{-\infty}^{\infty}df\frac{hf\omega_{\rm J}\exp(i2\pi f(\tau-\tau'))}{RI_0^2(\exp(hf/k_{\rm B}T)-1)}\,,
\label{eq:thermal}
\end{aligned}
\end{equation}
for the thermal noise described by Planck blackbody radiation law. The parentheses $\langle x\rangle$ denote the statistical average of the noise current, $f$, $T$, $h$, and $k_{\rm B}$ are radiation frequency,  temperature, Planck-, and Boltzmann constants, respectively. 
Note that, here the thermal noise described by the standard Planck spectrum~\eqref{eq:thermal}, rather than its high-temperature limit (i.e., the Nyquist noise: $\langle i_{\rm n} (\tau)i_{\rm n}(\tau')\rangle=4k_{\rm B}T/R$) used typically in Refs.~\cite{JJ3,thermal,Nyquist1,book}, as the present device works practically at the ultra-low temperature regime wherein the usual high-temperature approximation $k_{\rm B}T>>hf$ is not satisfied. 

\begin{figure}[htbp]
	\centering
	\subfigure[]{	\includegraphics[width=0.45\linewidth]{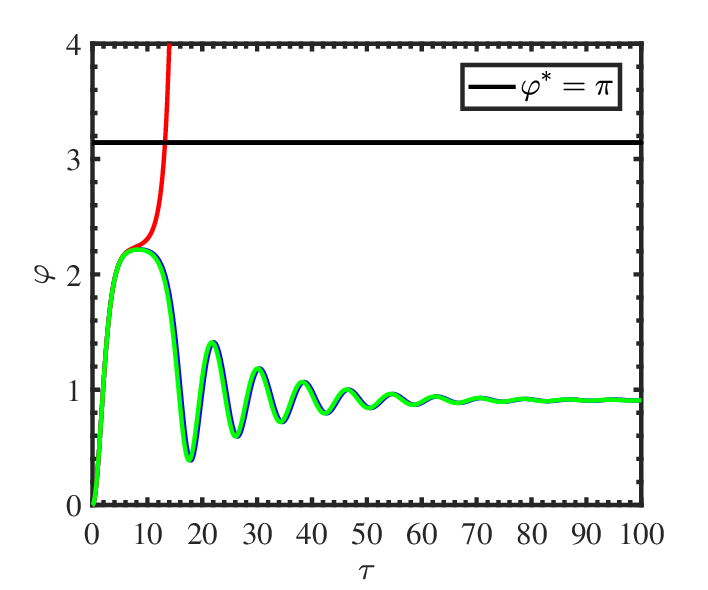}
	}
	\subfigure[]{     
    \includegraphics[width=0.45\linewidth]{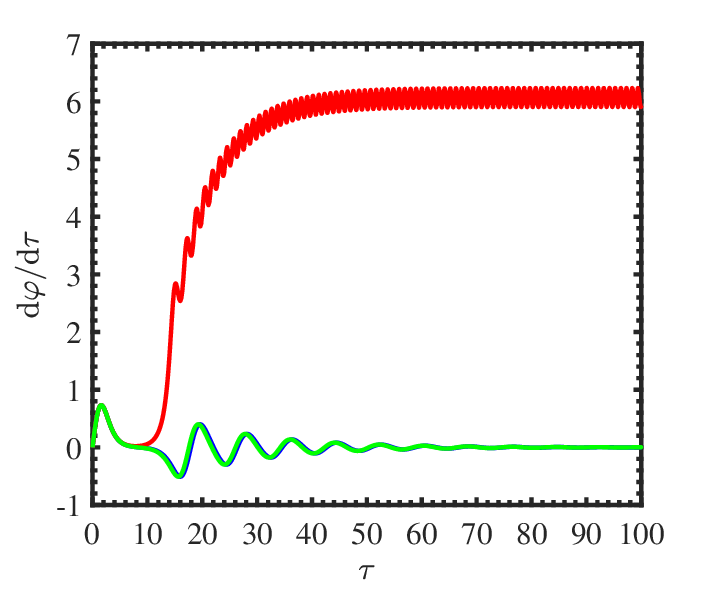}
    }
	\caption{The simulated phase dynamics (a) and voltage changes (b), by numerically solving  Eq.~\eqref{eq:Langevin}. Here, the length of the time step is set as $\delta t=0.0001$ and the JJ parameters are set as ~\cite{JJ-Al1}; $I_{\rm 0}=8.586~\upmu$A, $R=29~\Omega$, and $C=2700$~fF, respectively. The biased current is set as $i_{\rm b}=0.789$.}\label{F4}
\end{figure}

We use the Eulerian algorithm~\cite{Euler1,Euler2} to solve the above Langevin equation numerically to simulate the kinetic behavior of the ``phase particle''. For simplicity, the amplitude of the noise current is treated specifically as
\begin{equation}
	\begin{aligned}
		\Delta i_{\rm n}=\sqrt{\langle i_{\rm n}(\tau)i_{\rm n}(\tau')\rangle \delta t}{\rm N(0,1)}\,,\label{eq:noise}
	\end{aligned}
\end{equation}
where $\delta t$ is the time step, and $\rm N(0,1)$ denotes a Gaussian-distributed random number generated by using the Box-M\"{u}ller method~\cite{Box,noise2}. 
According to Eqs.~\eqref{eq:potential} and~\eqref{eq:high}, we know that the proper selection of bias current is particularly important for CBJJ to detect microwave weak signals. Too small a bias current requires a strong external signal to make the JJ be switched into the finite voltage state, and thus it is more difficult to detect the weak microwave signal. On the other hand, however, if the bias current is too large, the existing noise (rather than the signal) makes the JJ be directly switched into its finite voltage state, yielding the detection of the external input signal is fail. 
To look for the optimal bias current of the CBJJ for the sensitive detection of significantly weak microwave signal, we must make the relationship between the bias current $i_{\rm b}$ and switching probability by measuring the switched time of the CBJJ. This can be performed as follows. First, solving the Eq.~\eqref{eq:Langevin} under a certain given bias current, the switching time could be defined as the duration of the phase is evolved from the value $\varphi(0)$ to the time when $\varphi(\tau)>\varphi^{*}=\pi$, shown schematically as the red line in Fig.~\ref{F4}, wherein the green and blue lines refer to the condition $\varphi(\tau)>\varphi^{*}=\pi$ is not satisfied and thus the CBJJ is not switched.
Secondly, by repeatedly solving the Eq.~\eqref{eq:Langevin} with a common current $i_{\rm b}$ multiple times for the random initial phase $\varphi_i(0),\,i=1,2,...$, we get the switching times $P(n)$; Finally, we scan the bias current $i_{\rm b}$ from 0 to 1 and then find the relationship between the applied bias current $i_{\rm b}$ and the switching times $P(n)$. 
\begin{figure}[htbp]
	\centering
	\subfigure[]{		\includegraphics[width=0.45\linewidth]{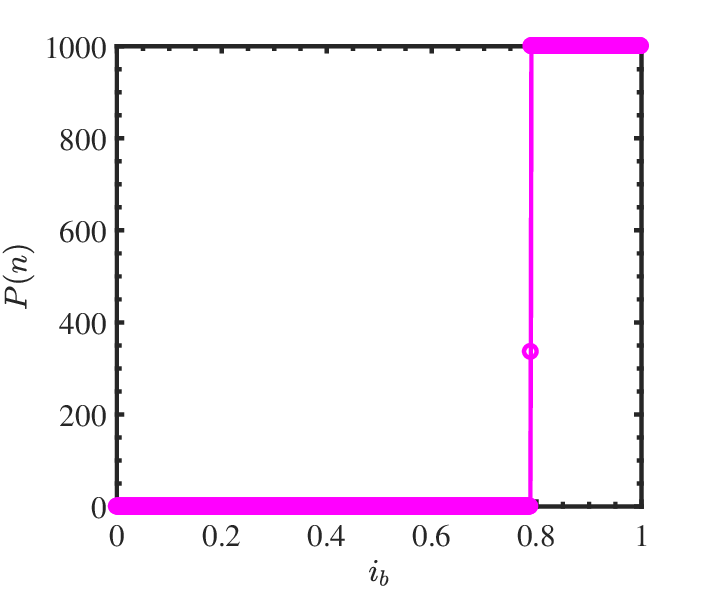}
	}
	\subfigure[]{		\includegraphics[width=0.45\linewidth]{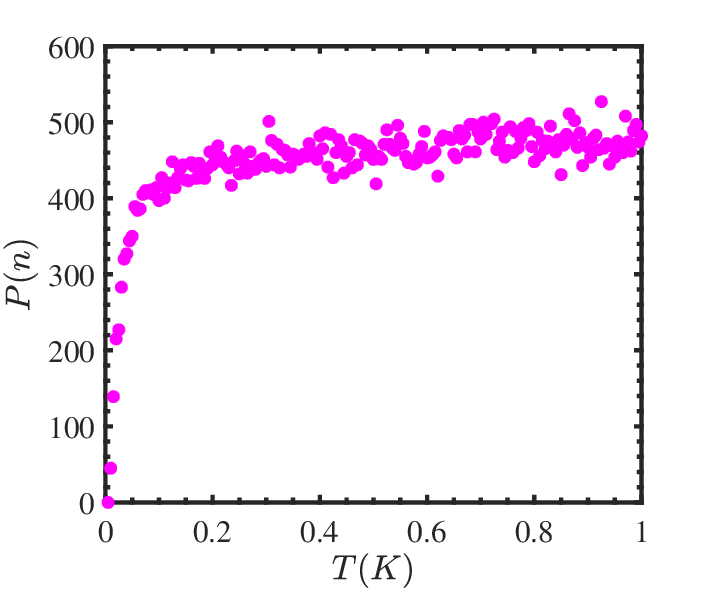}
    }
	\caption{The numerical simulation of the voltage-state switching of the CBJJ; (a) versus the normalized biased current $i_{\rm b}$ at $T=50$~mK, and (b) change with the temperature $T (K)$ for $i_{\rm b}=0.789$, respectively. Here, $P(n)$ refers to the switching times among the $1000$- times simulations. The time step and junction parameters are set as $\delta t=0.0001$, $I_{\rm 0}=8.586~\upmu$A, $R=29~\Omega$, and $C=2700$~fF, respectively.}\label{F5}
\end{figure}
From the simulated result shown in Fig.~\ref{F5}(a), one can see that, for the fixed background temperature, the switching times $P(n)$ are obviously related to the bias current; the smaller biased current refer to the fewer times of the CBJJ switches, and the optimal bias current for the CBJJ is estimated at $i_{\rm b}=0.789$. Also, for a given bias current, Fig.~\ref{F5}(b) shows that the switching times $P(n)$ of the CBJJ are increased with the increase of the temperature. This indicates that the dark switching count is strongly influenced by the temperature, and thus the lower temperature is the more beneficial to improve the sensitivity of the weak signal detection.

Based on the above simulation of the noise-induced switches, we now investigate how the single-induced switches could be distinguishably detected. Under the signal driving, the phase dynamics for the CBJJ can be described by the following equation,
\begin{equation}
\begin{aligned}
\frac{{\rm d}^2\varphi}{{\rm d}\tau^2}+\beta\frac{{\rm d}\varphi}{{\rm d}\tau}+\sin(\varphi)=i_{\rm b}+i_{\rm n}+i_{\rm s}.\label{eq:Langevin1}
\end{aligned}
\end{equation}
According to Eqs.~\eqref{eq:potential} and~\eqref{eq:high}, we know that the applied signal makes the depth of the potential well for trapping the phase particle become shallow, yielding the particle more easily escaped out. This implies that the switching time $\tau$, i.e., the time for the phase to be evolved from $\varphi(0)$ into the value  $\varphi(\tau)>\varphi^{*}=\pi$, might be lowered. Therefore, by comparing the difference between the switching time of the phase particle with and without the microwave signal driving, one can achieve the desired weak microwave signal detection. Certainly, the distinguishability of such a difference, i.e., the detection sensitivity is basically determined by the always-on noise. To quantify such a difference, the so-called Kumar Caroll (KC) index~\cite{d-kc1,d-kc2};
\begin{equation}
\begin{aligned}
d_{\rm KC}=\frac{|\langle \gamma_0\rangle-\langle \gamma_{1}\rangle|}{\sqrt{\frac{1}{2}(\sigma_{\gamma_0}+\sigma_{\gamma_1})}}\,,\label{eq:KC}
\end{aligned}
\end{equation}
could be used to represent the difference between two statistical distributions obtained of the phase particle with and without the signal driving. Above, $\gamma_0$ and $\gamma_1$ are the switching times of the phase particle in the CBJJ without and with the microwave signal driving, respectively. In the present work, they are simulated by numerically solving Eq.~\eqref{eq:Langevin} (without microwave signal) and Eq.~\eqref{eq:Langevin1} (with microwave signal) respectively for N times with the randomly initial phases. Certainly, due to the always-on thermal noise, the switching times $\tau_k^i, i=1,...,N$ must reveal certain   statistical distributions with the average 
\begin{equation}
\begin{aligned}
\langle \gamma_i\rangle=\frac{1}{N}\sum_{k=1}^{N}\tau_{k}^i\,,i=0,1\,,
\end{aligned}
\end{equation}
and the standard variance 
\begin{equation}
\begin{aligned}
\sigma_{\gamma_i}=\frac{1}{N(N-1)}\sum_{k=1}^{N}(\tau_{k}^i-\langle \gamma_i\rangle)^2\,,i=0,1\,.
\end{aligned}
\end{equation}
for the measurements of N times. Certainly, the larger the value of $d_{\rm KC}$ refers to the higher distinguishability between the two statistical distributions. 

\begin{figure}[htbp]
	\centering
\includegraphics[width=0.7\linewidth]{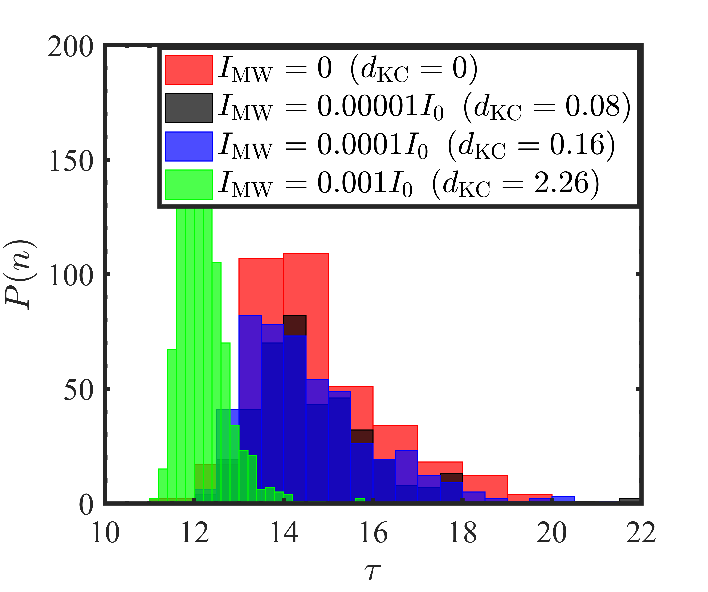}
	\caption{Simulation of the voltage-state switching times for the CBJJ driven by the different continuous wave microwave currents $I_{\rm MW}$. Here, the length of the time step is set as $\delta t=0.0001$ and the parameters CBJJ are set as~\cite{JJ-Al1}; $I_{\rm 0}=8.586~\mu$A, $R=29~\Omega$ and $C=2700$~fF, respectively. Also, the bias current, system temperature, and microwave signal frequency are set as: $i_{\rm b}=0.789$, $T=50$~mK, and $\omega_{\rm s}/2\pi\approx12.25$~GHz, respectively. The $d_{\rm KC}$ indexes of the statistical distribution of the switching events, compared to those with $I_{\rm MW}$ are calculated as $d_{\rm KC}=0, 0.08, 0.16$, and $2.26$, respectively.}\label{F6}
\end{figure}

With the typical JJ parameters~\cite{JJ-Al1}: $I_{\rm 0}=8.586~\upmu$A, $R=29~\Omega$, $C=2700$~fF, we repeatedly solve the Eq.~\eqref{eq:Langevin1} for $1000$ times and get the statistical distribution shown in Fig.~\ref{F6}, of the switching times, for different microwave drivings. 
Accordingly, the relevant KC indexes can be calculated to measure the distinguishing abilities between the statistical distributions with microwave drivings and those obtained for pure thermal noise. Specifically, one can see that if the amplitude of the applied microwave driving is sufficiently large, typically such as $I_{\rm MW}=0.001I_0$, the statistical distribution for the switching times caused by the microwave driving is reliably distinguished from that without the driving, i.e., $d_{\rm KC}=2.26>1$.
An open question is, how weak the microwave signal causes the statistical distribution of the switching times can be distinguished from those purely due to the background noise. Certainly, $d_{\rm KC}>1$ refers to the reliable discrimination~\cite{d-kc3}, but it does not mean that two distributions must be indistinguishable for $d_{\rm KC}<1$. Physically, there must be a minimum KC index (which is usually less than 1), i.e., two distributions are still discriminated, if $d_{\rm KC}>\min[d_{\rm KC}]<1$. This is because the detectability of an energy signal is basically determined by the noise-equivalent power (NEP) of the detector. On the other hand, $\min[d_{\rm KC}]$ refers to the distinguishing ability of the detectable power in the noise background. Therefore, the following task is to determine the $\min[d_{\rm KC}]$ by measuring the NEP of the detector.    

\begin{figure}[htbp]
	\centering
	\subfigure[]{	\includegraphics[width=0.45\linewidth]{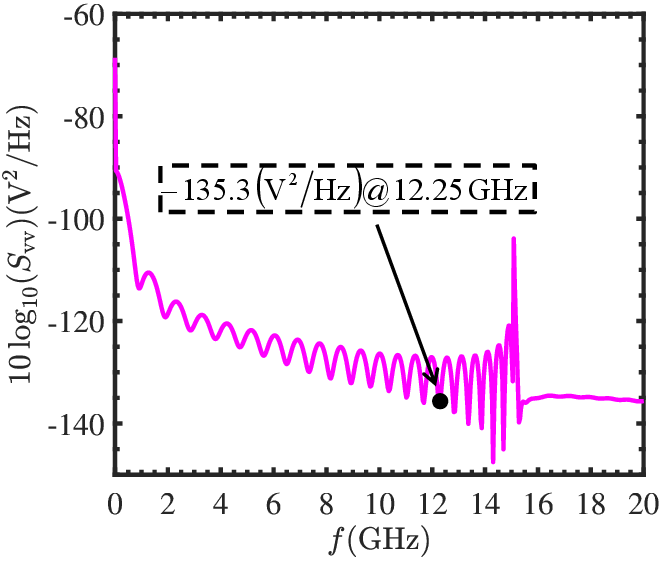}
	}
	\subfigure[]{	\includegraphics[width=0.45\linewidth]{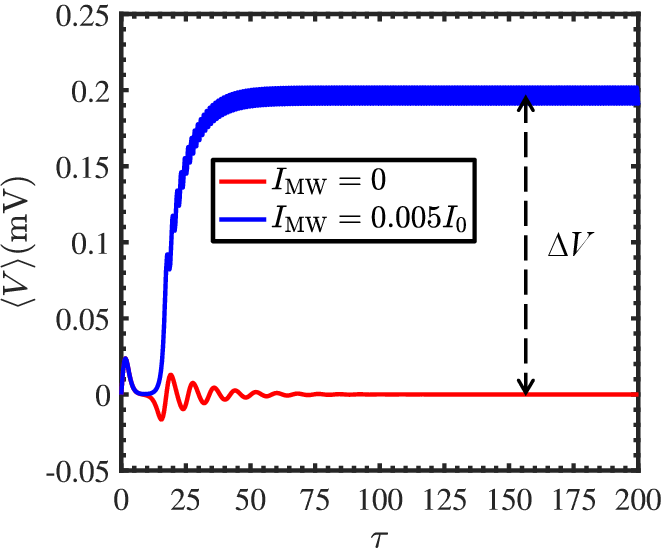}
    }
	\caption{(a) Numerical results for the thermal noise voltage spectrum of the CBJJ. Here, the length of the time step is set as $\delta t=0.0001$ and the CBJJ parameters are set as~\cite{JJ-Al1}: $I_0=8.586~\upmu$A, $R=29~\Omega$, $C=2700$~fF, and $i_{\rm b}=0.789$, respectively. Specifically, at $12.25$ GHz, we find that $10\log_{\rm 10}(S_{\rm vv})=-135.3 (\rm {V^2/Hz})$. (b) Numerical solution to the Eq.~\eqref{eq:Langevin1} for the time-average voltage of the CBJJ, without microwaves (red) and with microwaves driving $I_{\rm MW}=0.005I_{\rm0}$. Here, the normalized bias current of the CBJJ is set as $i_{\rm b}=0.789$. The difference between the two voltage averages refers to the responsivity of the CBJJ, i.e., $\Delta V\approx0.195$~mV.}\label{F7}
\end{figure}

\begin{figure}[htbp]
	\centering
    \includegraphics[width=0.7\linewidth]{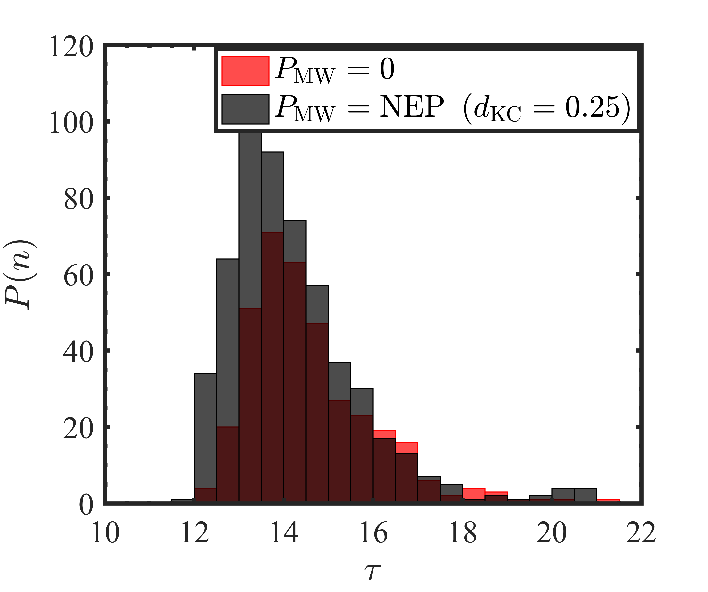}
	\caption{The simulated statistical distributions of the voltage switching times, by solving Eqs.~\eqref{eq:Langevin} and \eqref{eq:Langevin1} for $1000$ times, for calculating the NEP-corresponding $d_{\rm KC}$, i.e., the minimum KC index of the CBJJ. Here, the length of the time step is set as $\delta t=0.0001$ and the CBJJ parameters are set as~\cite{JJ-Al1}: $I_{\rm 0}=8.586~\upmu$A, $R=29~\Omega$, and $C=2700$~fF, respectively. The bias current, background temperature, and microwave signal frequency are set as $i_{\rm b}=0.789$, $T=50$~mK and $\omega_{\rm s}/2\pi\approx12.25$~GHz, respectively.}\label{F8}
\end{figure}

Practically, the NEP of the detector can be calculated as~\cite{NEP}
\begin{equation}
\begin{aligned}
{\rm NEP}=\frac{S_{\rm v}}{S}\,,
\end{aligned}
\end{equation}
where $S_{\rm v}(\rm {V/\sqrt{Hz}})$  and $S=\Delta V/P_{\rm in}(\rm {V/W})$ are the noise voltage and the detection sensitivity, respectively. $P_{\rm in}=I_{\rm MW}^2R$ is the input signal power, and $\Delta V$ is the relevant response voltage.
For the present CBJJ device, its NEP can be determined as follows. First, the voltage power spectral density $S_{\rm vv}$ of the device can be obtained by solving the Eq.~\eqref{eq:Langevin} and shown specifically in Fig.~\ref{F7}(a).
Consequently, the noise voltage, typically for the signal of frequency $12.25$ GHz, can be calculated as $S_{\rm v}=\sqrt{S_{\rm vv}}=1.7179\times10^{-7}(\rm {V/\sqrt{Hz}})$. Next, the response voltage $\Delta V$ of the device can be determined by comparing the difference of the averaged voltages with and without the signal input. With the relevant numerical simulation shown in Fig.~\ref{F7}(b), we find $\Delta V\approx0.195$~mV for the input $I_{\rm MW}=0.005I_0$.
Thirdly, the NEP of the CBJJ detector, with the typical parameters $I_{\rm 0}=8.586~\upmu$A, $R=29~\Omega$, $C=2700$~fF and $i_{\rm b}=0.789$, is calculated as $S=\Delta V/P_{\rm in}=3.6485\times10^{9}$~(V/W), ${\rm NEP=4.7085\times10^{-17}~W/\sqrt{Hz}}$ for the detection of the signal at the frequency 12.25~GHz. 
Accordingly, with the NEP estimated above, the current of the detectable weakest signal is estimated as $I_{\rm min}=\sqrt{{\rm NEP}/R}=1.2742$~nA. In Fig.~\ref{F8}, we simulate the probabilistic distributions of the phase particle escapes with and without the input of the signal current $I_{\rm min}$. The relevant KC index, i.e., the minimum KC index, is calculated as $\min[d_{\rm KC}]=0.25$. 
This means that, with the present CBJJ device, the signals that make $d_{\rm KC}>\min[d_{\rm KC}]=0.25$ should be detectable. 
%

\section{Estimating the detectable microwave photon number of the CBJJ detector}
In Sec.~II, we provided an effective method to simulate the escape events of a driven phase particle in the presence of thermal noise. A minimum KC index $\min[d_{\rm KC}]$ is introduced to describe the detectability of the signal current along the CBJJ device. Certainly, such a detectability is strongly dependent on the physical parameters of the CBJJ device. Tab.~\ref{tab1} lists the relevant $\min[d_{\rm KC}]$-parameter for the typical CBJJ detectors.  
\begin{table}[htbp]
\setlength{\tabcolsep}{8mm}{
  \centering
  \caption{The calculated $\min[d_{\rm KC}]$ indexes for different CBJJs~\cite{JJ-Al1,JJ-Al2,our} under the best bias current and resonance frequency point. Here, the time step and critical phase are set as $\delta t=0.0001$ and $\varphi^{*}=\pi$, respectively.}\label{tab1}
   \resizebox{\linewidth}{!}{
   \begin{tabular}{cccccc}
    \toprule[0.8pt]
                                      & JJ1         & JJ2         & JJ3       \\
    \specialrule{0.5pt}{0pt}{0pt}
    $I_0~(\upmu$A)                    & $8.586$       & $2$         & $0.975$         \\
    $R~(\Omega)$                      & $29$          & $130$        & $290$      \\
    $C$~(fF)                          & $2700$        & $630$        & $93$       \\
    $T$~(mK)                          & $50$          & $48$         & $50$        \\
    $i_{\rm b}$                               & $0.789$   & $0.786$  & $0.825$   \\
    $I_{\rm MW}$                               & $0.005I_{\rm 0}$   & $0.005I_{\rm 0}$  & $0.005I_{\rm 0}$   \\
    $\omega_{\rm s}/2\pi$~(GHz)              & $12.25$   & $12.28$  & $21.37$   \\
    ${\rm NEP(aW/\sqrt{Hz})}$                      & $47.085$   & $26.848$  & $9.623$       \\
    $\min[d_{\rm KC}]$                      & $0.25$   & $0.32$  & $0.17$       \\
    \specialrule{0.8pt}{0pt}{\belowrulesep}
  \end{tabular}
   }}
\end{table}
Obviously, the lower $\min[d_{\rm KC}]$-parameter refers to the lower NEP of the CBJJ detector, and thus the higher sensitivity for the detection of microwave photons.

\begin{figure}[htbp]
	\includegraphics[width=0.7\linewidth]{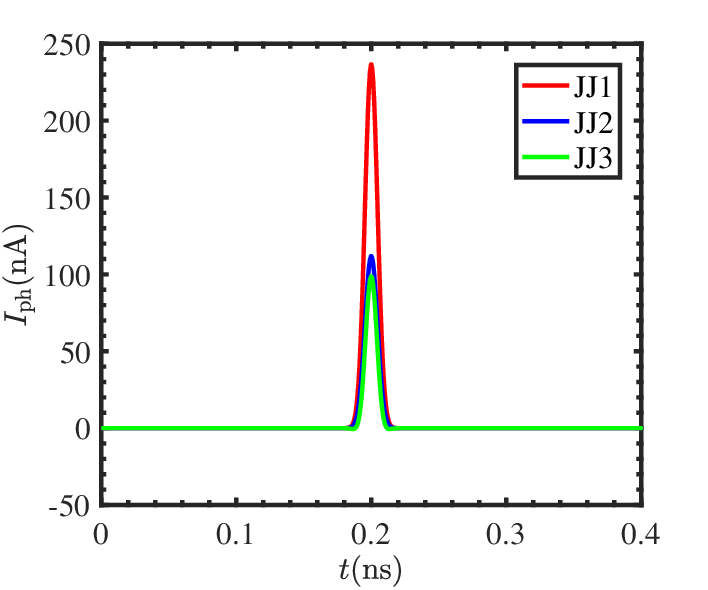}
	\centering
	\caption{The shapes of the single-photon microwave current pulses through the three JJs listed in Tab.~\ref{tab1} with different resistances. The pulse width $t_{\rm ph}$ and arrival time $t_{\rm d}$ are set as $0.005$ ns and $0.02$ ns, respectively.}\label{F9}
\end{figure}

It is emphasized that, in Sec.~II the microwave signal is treated as a coherent continuous-wave signal. Below, we assume that a microwave current pulse~\cite{Nyquist2}
\begin{equation}
\begin{aligned}
I_{\rm ph}(\omega_{\rm ph}, t_{\rm ph}, t)&=\sqrt{N}\sqrt{\frac{\hbar\omega_{\rm ph}}{Rt_{\rm ph}}}\exp\left(-\frac{1}{2}(\frac{t-t_{\rm d}}{t_{\rm ph}})^2\right)\\
&\times\cos(\omega_{\rm ph}(t-t_{\rm d}))\,.\label{eq:ph}
\end{aligned}
\end{equation}
with $N$, $\omega_{\rm ph}$, $t_{\rm ph}$ and $t_{\rm d}$ being respectively its containing number of photons, frequency, width, and arrival time, etc., flow through the CBJJ detector with the resistance $R$. The shapes of these pulses through the above three CBJJs are shown schematically in Fig.~\ref{F9}.
For these cases, the dynamics for the phase particles of the CBJJ devices should be changed as  
\begin{equation}
\begin{aligned}
\frac{{\rm d}^2\varphi}{{\rm d}\tau^2}+\beta\frac{{\rm d}\varphi}{{\rm d}\tau}+\sin(\varphi)=i_{\rm b}+i_{\rm n}+i_{\rm ph}(\omega'_{\rm ph}, \tau_{\rm ph}, \tau).\label{eq:Langevin2}
\end{aligned}
\end{equation}
from Eq.~\eqref{eq:ph}, where the dimensionless pulse current, microwave frequency, and pulse width are respectively normalized to the critical current $I_0$ and the Plasma frequency $\omega_{\rm J}$ of the CBJJ, i.e., $i_{\rm ph}=I_{\rm ph}/I_0$, $\omega'_{\rm ph}=\omega_{\rm ph}/\omega_{\rm J}$ and $\tau_{\rm ph}=t _{\rm ph}\omega_{\rm J}$. 

\begin{figure*}[htbp]
	\centering
	\subfigure[]{	\includegraphics[width=0.3\linewidth]{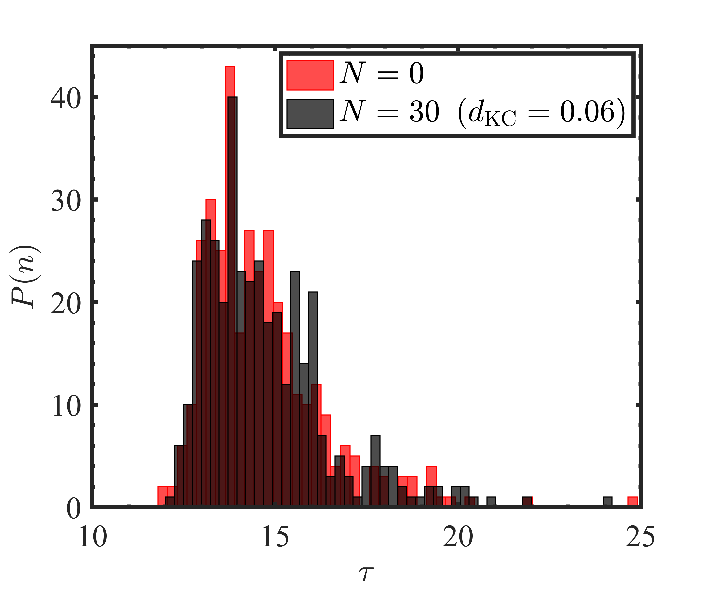}
	}
	\subfigure[]{	\includegraphics[width=0.3\linewidth]{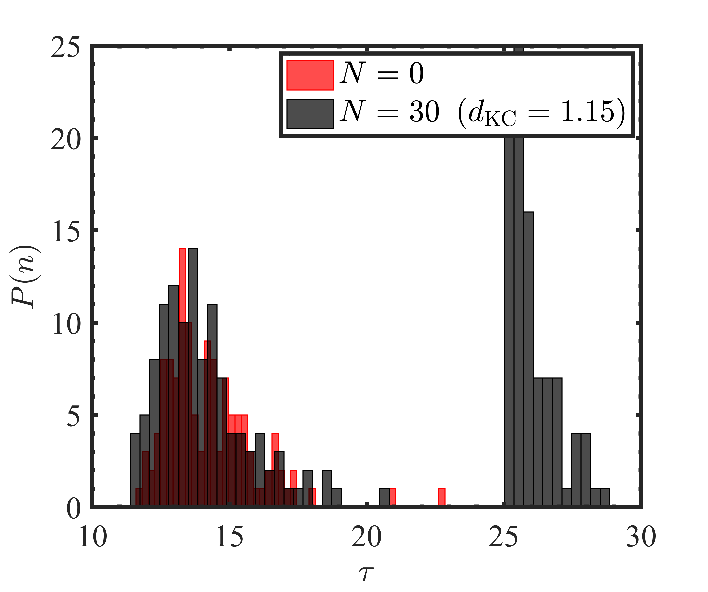}
    }
    \subfigure[]{	\includegraphics[width=0.3\linewidth]{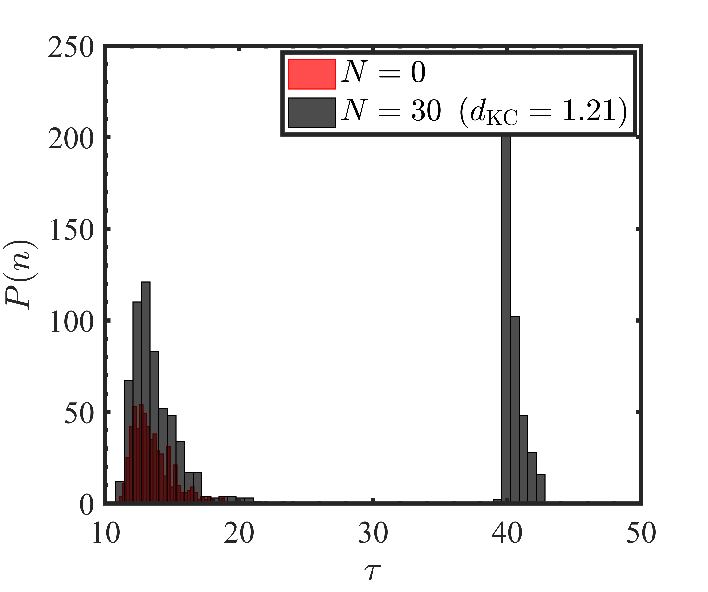}
    }
	\caption{Numerical simulation results for the voltage state switching times and the corresponding KC-indices for the JJ1 (a), JJ2 (b), and JJ3 (c), driven by the microwave pulses with $N=0, 30$-photons. They are obtained by solving Eqs.~\eqref {eq:Langevin} and \eqref{eq:Langevin2} for 1000 times, respectively. Here, the time step is set as $\delta t=0.0001$.}\label{F10}
\end{figure*}

For the typical three CBJJs listed in Tab.~\ref{tab1}, we numerical solve the Eq.~\eqref{eq:Langevin2} to simulate the relevant escape events of the phase particle without and with the pulse driving containing, e.g., 30-microwave photons, respectively. The simulated results are shown specifically in Fig.~\ref{F10}. Given the $\min[d_{\rm KC}]$-parameter, similar to the usual NEP one, depends completely on the parameters (i.e., the capacitance $C$, resistance $R$, and critical current $I_0$ of the JJ and the biased current $i_{\rm b}$) of the CBJJ device, the detectability of the applied microwave current pulse is thus dependent of if the measured $d_{\rm KC}$ is larger than the corresponding $\min[d_{\rm KC}]$-parameter, or not.
\begin{figure*}[htbp]
	\centering
	\subfigure[]{	\includegraphics[width=0.3\linewidth]{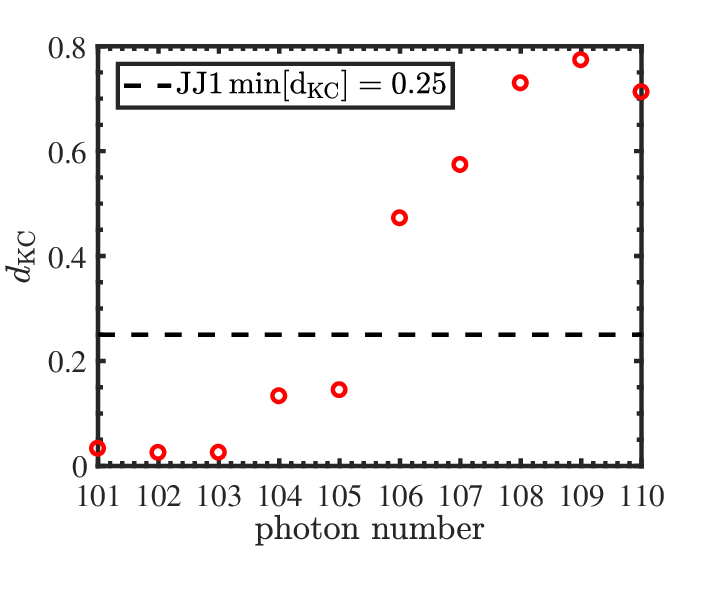}
	}
	\subfigure[]{	\includegraphics[width=0.3\linewidth]{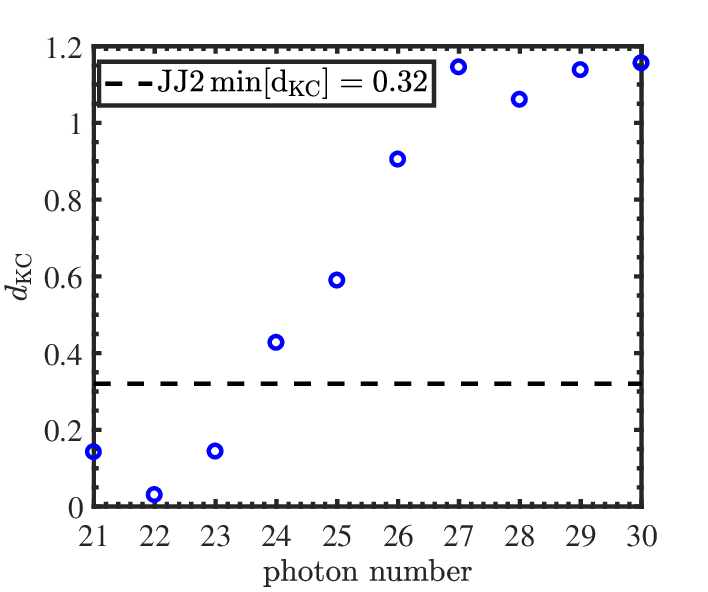}
    }
    \subfigure[]{	\includegraphics[width=0.3\linewidth]{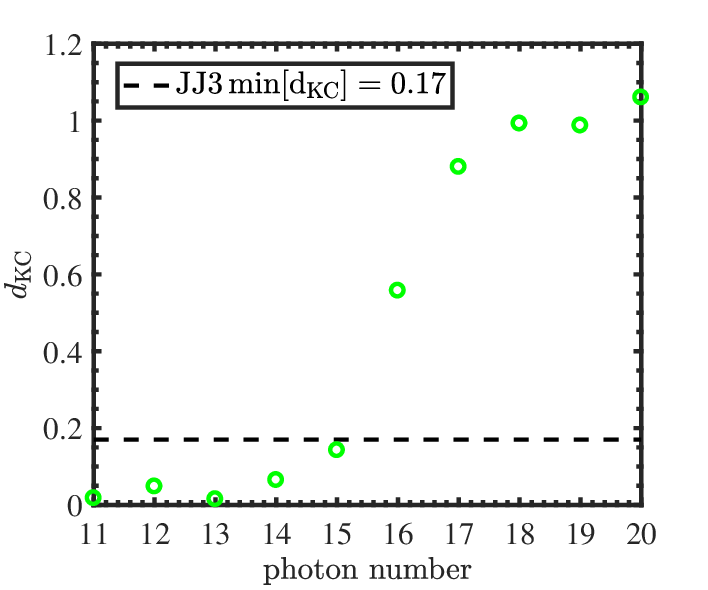}
    }
	\caption{Numerical predictions of the detectable minimum number of the microwave photons, with the three JJs listed in Tab.~\ref{tab1}. Here, the time step is set again as $\delta t=0.0001$. It is seen that the minimum number of microwave photons by using the JJ1, JJ2, and JJ3 are $106$ (a), $24$ (b), and $16$ (c), respectively.}\label{F11}
\end{figure*}
Therefore, the simulated results shown in Fig.~\ref{F10} indicate that the applied microwave current pulse of the frequency $\omega_{\rm ph}=12.25$ GHz and containing 30-photons, should not be detected by the JJ1-device, as the measured value ($d_{\rm KC}$=0.06) is manifestly less than its $\min[d_{\rm KC}]$-parameter value $0.25$. While such a microwave current pulse could be detected respectively by using the JJ2- and JJ3 devices, as the measured values of the $d_{\rm KC}$ indexes (i.e., $1.15$ and $1.21$, shown respectively in the figure) are manifestly larger than the corresponding $\min[d_{\rm KC}]$-parameter values (i.e., $0.25$ and $0.17$, listed respectively in Tab.~\ref{tab1}.)
This implies that the detectable photon number is basically determined by the $\rm min[d_{KC}]$-parameter of the CBJJ detector. In Tab.~II, we list the minimum number of the detectable microwave photons of the frequency $12.25$ by using the CBJJ detectors listed in Tab.~\ref{tab1}. 
\begin{table}[h]
\setlength{\tabcolsep}{8mm}{
  \centering
  \caption{The calculated $\min[d_{\rm KC}]$ indexes and detectable photons N for different CBJJs~\cite{JJ-Al1,JJ-Al2,our} under the best bias current and resonance frequency point. The time step is $\delta t=0.0001$.}\label{tab2}
   \resizebox{\linewidth}{!}{
   \begin{tabular}{cccccc}
    \toprule[0.8pt]
                                      & JJ1        & JJ2       & JJ3       \\
    \specialrule{0.5pt}{0pt}{0pt}
    $i_{\rm b}$                              & 0.789   & 0.786  & 0.825   \\
    $\min[d_{\rm KC}]$                      & $0.25$   & $0.32$  & $0.17$       \\
    $\rm N$                               & $104$   & $24$  & $21$   \\
    \specialrule{0.8pt}{0pt}{\belowrulesep}
  \end{tabular}
   }}
\end{table}
Obviously, the detector with the lower minimum $d_{\rm KC}$-index possesses the higher detection sensitivity. Given the resistance, capacitance and critical current of the Josephson junction are uncontrollable (once it was fabricated), increasing the bias current seems to be the only effective approach to enhance the detection sensitivity of the CBJJ detector. Based on this idea, in Fig.~\ref{F11} we numerically predict how the minimum number of microwave photons can be detected by using the JJs listed in Tab.~\ref{tab1} by optimizing the relevant bias current for the resonant detection of the microwave photons. Here, the frequency of the applied microwave photon is $\omega_{\rm s}\cong\omega_{\rm J}^{*}=(1-i_{\rm b}^2)^{1/4}\omega_{\rm J}$. It is seen that the minimum detectable number of the resonant photons by using the JJ1 device, with the optimized bias current, is predicted as 106 photons with the frequency being 12.25 GHz; While, the detectable minimum numbers of the resonant photons by using the JJ2- and JJ3 detectors are predicted as 24 at 12.28 GHz and 16 at 21.37 GHz, respectively. They are the approached EQLDs. Probably, to implement the desired EQLD, i.e., the single-photon detection, is required to increase the resistance of the CBJJ and optimize the biased current, e.g., by alternatively using the ac biases. 

\section{Conclusions and Discussions}
In summary, by numerically simulating the escaped dynamics for the driven phase particle, we analyzed the detectability of the weak microwave current by using the CBJJ detector. We showed that the difference between the statistical distributions of the probabilistically escaped events of the phase particle, driven by the thermal noises and the applied microwave current, could be quantified by the $d_{\rm KC}$-index. Physically, the NEP of a detector refers to its $\min[d_{\rm KC}]$-index. Therefore, if the measured $d_{\rm KC}$-index is larger than the $\min[d_{\rm KC}]$-index of the detector, the applied microwave current signal is detectable. For the typical CBJJ devices, we have predicted their detectable minimum number of resonant microwave photons, by numerically characterizing the relevant values of the $\min[d_{\rm KC}]$-index, which practically describes the abilities of the devices to discriminate the weakest signals from the noise background. Thanks to the very low operating temperature of the CBJJ detector, the significantly weak microwave current pulse signals, e.g., lowered to a dozen microwave photons approaching to the quantum limit, are still detectable. 

It is worth emphasizing that, in our numerical analysis, the NEPs of the typical CBJJ detectors, refers to the measurable minimum $d_{\rm KC}$-indexes, were estimated as at $\sim 10^{-18}~{\rm W/\sqrt{Hz}}$ level. Following Ref.~\cite{NEP}, however, if the noise voltage $S_{\rm v}$ is reduced to $0.1~{\rm nV/\sqrt{Hz}}$ by using the lock-in technique, the NEP of these CBJJ detectors could be boosted to the $10^{-23}~{\rm W/\sqrt{Hz}}$ order of the magnitude. This implies that, besides the various developing quantum measurements~\cite{quam-JJ1,quam-JJ2,quam-JJ3}, it is still possible, at least theoretically, to implement the desired QLED of microwave photon by using the optimized CBJJ detector~\cite{our2}.

\end{document}